\documentclass[aps,prl,reprint,superscriptaddress,nofootinbib]{revtex4-1}
\usepackage{amsmath}
\usepackage{amssymb}
\usepackage{graphicx}
\usepackage[caption=false]{subfig}
\usepackage{enumitem}
\usepackage{float}
\usepackage{color}
\usepackage{comment}
\usepackage{mathtools}
\usepackage[percent]{overpic}
\usepackage{bm}
\usepackage{ulem}
\usepackage{rotating}
\usepackage{lipsum}
\usepackage{blindtext}
\usepackage{cancel}
\usepackage{hyperref}
\usepackage{tikz}   
\usepackage{quantikz}
\usetikzlibrary{quantikz} 
\usepackage[T1]{fontenc} 
\usepackage{chemfig} 
\usepackage{booktabs} 
\newcommand{\tr}{\text{Tr}}

\definecolor{ultramarineblue}{rgb}{0.25, 0.4, 0.96}

\begin{document}

\title{ Experimental Activation of Strong Local Passive States with Quantum Information}

%
%
\author{Nayeli A. Rodr\'{i}guez-Briones}
\email[e-mail address: ]{nayelongue@berkeley.edu}
\affiliation{Department of Chemistry, University of California, Berkeley, California 94720, USA}
\affiliation{Miller Institute for Basic Research in Science, 468 Donner Lab, Berkeley, CA 94720, USA}
\author{Hemant Katiyar}
\affiliation{Institute for Quantum Computing, University of Waterloo, Waterloo, Ontario, N2L 3G1, Canada}
\affiliation{Department of Physics \& Astronomy, University of Waterloo, Waterloo, Ontario, N2L 3G1, Canada}

\author{Eduardo~Mart\'{i}n-Mart\'{i}nez}
\affiliation{Institute for Quantum Computing, University of Waterloo, Waterloo, Ontario, N2L 3G1, Canada}
\affiliation{Perimeter Institute for Theoretical Physics, 31 Caroline St. N., Waterloo, Ontario, N2L 2Y5, Canada}
\affiliation{Department of Applied Mathematics, University of Waterloo, Waterloo, Ontario, N2L 3G1, Canada}
 \author{Raymond Laflamme}
\affiliation{Institute for Quantum Computing, University of Waterloo, Waterloo, Ontario, N2L 3G1, Canada}
\affiliation{Department of Physics \& Astronomy, University of Waterloo, Waterloo, Ontario, N2L 3G1, Canada}
\affiliation{Perimeter Institute for Theoretical Physics, 31 Caroline St. N., Waterloo, Ontario, N2L 2Y5, Canada}

\begin{abstract}

Strong local passivity is a property of multipartite quantum systems from which it is impossible to extract energy locally. 
Surprisingly, if the strong local passive state displays entanglement, it could be possible to locally activate energy density by adding classical communication between different partitions of the system, through so-called ``quantum energy teleportation'' protocols. Here, we report both the first experimental observation of local activation of 
energy density on an entangled state and the first realization of a quantum energy teleportation protocol using nuclear magnetic resonance on a bipartite quantum system.

\end{abstract} 

\maketitle




\textit{Introduction.---}Methods to extract and transfer energy from physical systems at the quantum scale have been developed recently using tools from quantum information processing and quantum thermodynamics~\cite{brunner2012virtual,linden2010small,levy2012quantum_refrigerator,boykin2002algorithmic,schulman1999molecular,fernandez2004algorithmic,schulman2005physical,schulman2007physical,sorensen:qc1990a,sorensen1991entropy,park2016heat,rodriguez2017correlation,PhysRevLett.116.170501,kose2019algorithmic,rodriguez2017heat,alhambra2019heat,niedenzu2018quantum,kieu2004second,quan2007quantum,cotler2019quantum,zaletel2021preparation,clivaz2019unifying,gluza2021quantum}.
%
%
But can these tools allow us to activate energy extraction from quantum systems in which outgoing energy flows are locally blocked~\cite{frey2014strong,alhambra2019fundamental}? Or to activate locally hidden energy in entangled ground states? One may be tempted to answer `no', as this would seem to involve activating zero-point energy, which is generally considered impossible. However, we will discuss that the answer to these questions is nuanced and that zero-point energy density can be activated using quantum informational tools.

\begin{figure}[ht]
\centering
\includegraphics[width=1\linewidth]{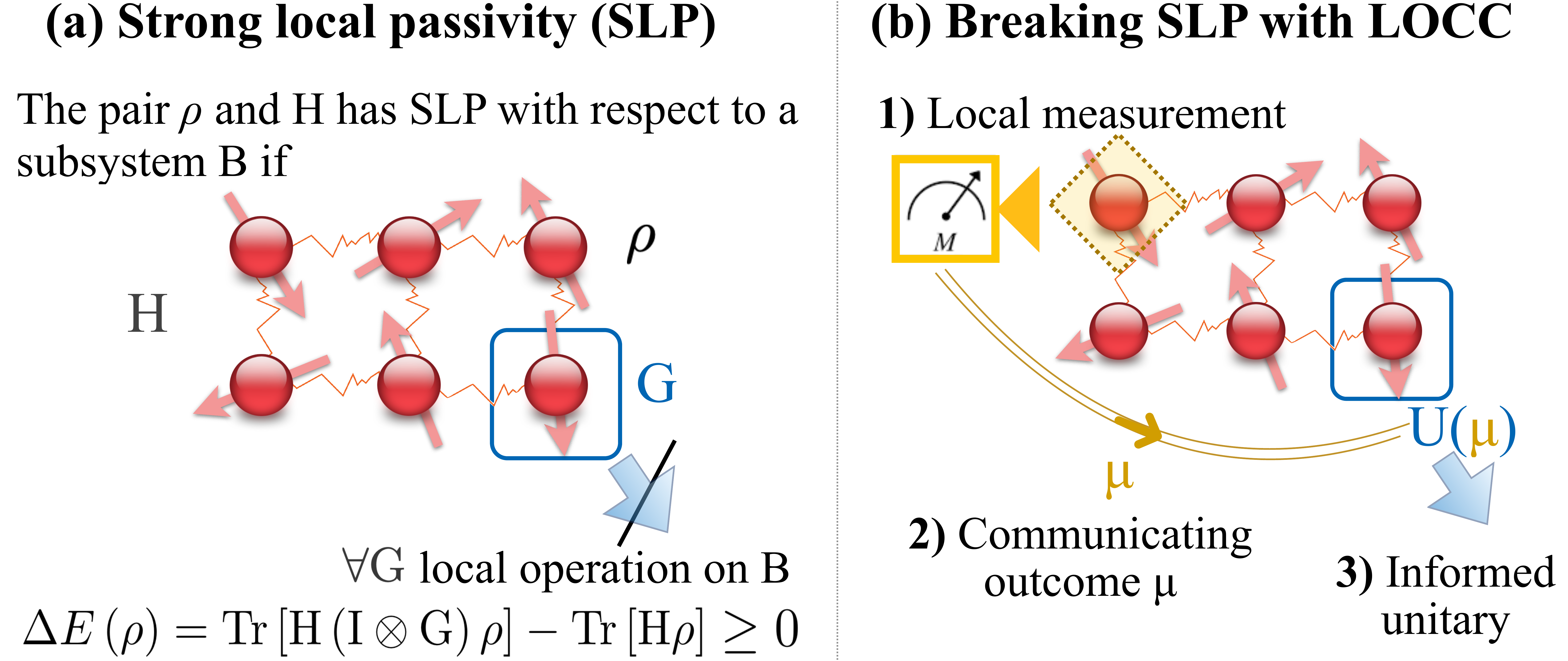}
\caption{\textbf{(a)} A quantum state $\rho$ is defined to be strong local passive with respect to a Hamiltonian $H$ and a subsystem if energy cannot be extracted through any direct local quantum operation G applied on the subsystem 
\textbf{(b)} Steps for breaking strong local passivity  by adding local operations and classical communication (LOCC).}
\label{fig:SLP_noSLPwithLOCC}
\end{figure}

The quantum states from which it is impossible to extract energy via general local access on a single subsystem receive the name of strong local passive (SLP) states~\cite{frey2014strong,alhambra2019fundamental,CPlocalPassivity} (Fig. \ref{fig:SLP_noSLPwithLOCC}). This distinctive property of strong local passivity is present in a wide range of states, from ground states to thermal states below a critical temperature and even in strongly coupled heat baths in the thermodynamic limit. The necessary and sufficient conditions for this property were presented in Ref. \cite{alhambra2019fundamental}.

Strong local passivity provides new insights into the emergent thermodynamic behavior arising from the interplay between entanglement and localization, such as understanding the allowed flows of energy and information within entangled quantum systems. Along these lines, a fundamental question is how and when strong local passivity can be broken. Certainly, finding methods to activate the non-directly available energy in SLP states can bring fascinating physical scenarios, such as activating entangled ground states. Indeed, in interacting multipartite quantum systems, the ground state can have positive (and negative) energy density regions due to its entanglement~\cite{zero-pointdensity}. However, the corresponding energy is not directly available since any action attempting to extract it directly will only give energy to the system. Could this energy be activated by driving the system differently?

This question was answered by Masahiro Hotta, who showed that, under certain conditions, it is possible to activate strong local passive states by allowing local operations and classical communication (LOCC) to exploit correlations between distant parts of the system~\cite{hotta2010controlled,frey2013quantum,hotta2014quantum,frey2014strong,hotta2009quantum,hotta2008protocol,trevison2015quantum,hotta2010energy,Verdon2016}. 
He introduced the family of protocols known under the general name of quantum energy teleportation (QET), which enables the activation of local energy using informed local operations that depend on the outcome of distant measurements on other sides of the system. Specifically, in a QET protocol, a local measurement is made on a subsystem (A) far from the subsystem (B) where the energy is blocked. Then, the outcome of this measurement is communicated to B's side. Because of the correlations, this outcome allows, to some extent, to predict and design an informed local operation to extract the previously inaccessible energy, see Fig.~\ref{fig:SLP_noSLPwithLOCC}. It is important to note that despite the name of the protocol, it does not imply that the energy injected during A's measurement disappears from A's surroundings to appear around B. 
Indeed, the key feature of the QET protocol is to ensure that the energy extracted comes only from the previously unavailable energy. Thus, for QET is crucial that the local measurement on A does not raise the energy in B's surroundings---which can be achieved by using measurement operators that commute the interacting Hamiltonian term---and that the protocol must be performed within a time shorter than the energy propagation time scale on the system.
Beyond the importance of QET in the activation of passive states, QET has also been suggested to be relevant to understanding a broad range of situations---from the black-hole information loss problem~\cite{hotta2010controlled,braunstein2007quantum,hayden2007black,hosoya2002quantum} and violations of energy conditions in quantum field theory in curved spacetimes~\cite{funai2017engineering} to technological applications such as local cooling of many-body systems with restricted measurements~\cite{rodriguez2017correlation}.

%
%
%
%

However, despite the potential applications of the QET protocol, it had not yet been realized in an experiment. Experimental proposals did exist, for instance, using a semiconductor exhibiting the quantum Hall effect~\cite{yusa2011quantum}, but no actual experiments had been conducted. This work presents the first experimental realization of a quantum energy teleportation protocol, demonstrating energy activation in a strong local passive state, in particular the local activation of zero-point energy density. The experiment was carried out using nuclear magnetic resonance on a system of three qubits in the ground state of an interacting simulated Hamiltonian. The experimental results show energy extraction from a system initially in an SLP state, beyond local and ambient noise, without energy transfer through the system. Our experiment demonstrates the feasibility of the control required for a QET protocol and the first evidence of activation of local zero-point energy density in an entangled ground state under experimental conditions. Furthermore, we present an optimized, fully unitary QET model and an analytical solution for the maximum extractable energy for our system.


%

We begin by summarizing the minimal QET model~\cite{hotta2010energy} to show how it is possible to break strong local passivity to activate regions of an entangled ground state. Then, we present the equivalent fully unitary version of QET implemented in the experiment, followed by the experimental results and conclusions.


\textit{Minimal QET model}.---Consider two interacting qubits, A and B, with a Hamiltonian that has a nondegenerate fully-entangled ground state. An example of such a Hamiltonian is
\begin{equation} 
\label{eq:Hamiltonian}
    H=H_\textsc{a}+H_\textsc{b}+V,
\end{equation}
${\rm with} \: H_\nu=-h_\nu\sigma^\nu_z + h_\nu f(h_\textsc{a},h_\textsc{b},\kappa) \openone, \: {\rm for} \: \nu\in\{\text{A},\text{B}\} \: {\rm and}$
%
%
\begin{equation} \label{eq:interaction}
V=2\kappa\sigma_x^\textsc{a}\sigma_x^\textsc{b}+\frac{4\kappa^2}{h_\textsc{a}+h_\textsc{b}} f(h_\textsc{a},h_\textsc{b},\kappa)\openone
\end{equation}
where $h_\textsc{a},\: h_\textsc{b}, \: {\rm and} \: \kappa$ are positive constants and the function
$f(h_\textsc{a},h_\textsc{b},\kappa)$ is chosen such that the ground state  $|g\rangle$ of the full Hamiltonian has vanishing expectation values for each of its terms $(\langle g|H_\textsc{a}|g\rangle=\langle g|H_\textsc{b}|g\rangle=\langle g|V|g\rangle=0)$ for convenience and without loss of generality. For this Hamiltonian,
$f(h_\textsc{a},h_\textsc{b},\kappa)=\left(4\kappa^2/(h_\textsc{a}+h_\textsc{b})^2+1\right)^{-\frac{1}{2}}$.
 
This Hamiltonian has a non-degenerate fully entangled ground state
 \begin{align} \label{eq:ground} 
  |g\rangle=\left(F_{+}\ket{00}_\textsc{ab}-F_{_-}\ket{11}_\textsc{ab}\right)/\sqrt{2},
 \end{align}
where 
%
$F_\pm=\sqrt{1\pm f(h_\textsc{a},h_\textsc{b},\kappa)}$, satisfying the sufficient conditions to have a family of SLP states (see Ref.~\cite{frey2014strong}). 
While the total Hamiltonian is a nonnegative operator ($H\geq0$ since its lowest eigenvalue is 0), $H_\textsc{b}\: {\rm and} \: H_\textsc{b}+V$ allow negative eigenvalues which could yield negative energy density in B's surroundings. A QET protocol can locally activate that energy, as described below.



%

Minimal QET model, implemented by ${\rm Alice}\: \&\: {\rm Bob}$:

\textbf{Step 1:} Alice measures subsystem A using a positive operator-valued measure (POVM) with measurement operators $M_\textsc{a}(\mu)$ that commute with the interacting Hamiltonian term $(\left[M_\textsc{a}(\mu),V\right]=0)$, to ensure that the energy injected during the measurement does not raise the energy of subsystem B~\cite{POVMnotraisingenergyB,hotta2010energy}.

\textbf{Step 2:} Alice communicates the measurement result $\mu$ to Bob in a time $t_\mu$ shorter than the coupling time scale $t_c\sim 1/\kappa$ to avoid the energy infused in A during the measurement propagates to B during that time.

\begin{figure}[ht]
\centering
\includegraphics[width=1\linewidth]{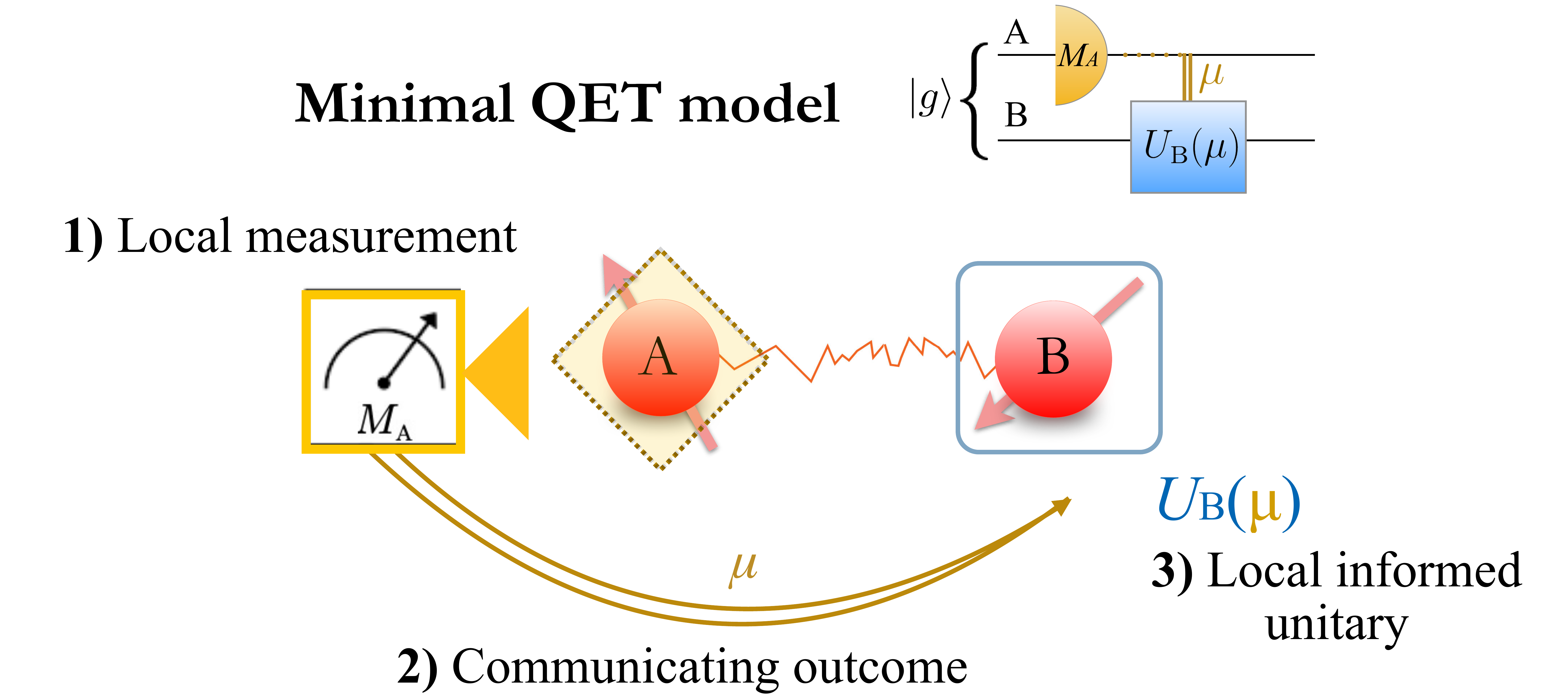}
\caption{{\color{black} Minimal QET protocol steps, performed on a pair of qubits initially in a strong local passive state:
\textbf{(1)} A local measurement is performed on qubit A using measurement operators $\{M_\textsc{a}\}$ that ensure the energy injected during the measurement remains local to A and does not raise the energy in B's surroundings. \textbf{(2)} The measurement outcome is communicated to B in a time shorter than the energy propagation time scale. \textbf{(3)} Based on this information, a local unitary operation is performed on B to extract energy.}}
\label{fig:QETprotocol}
\end{figure}

\textbf{Step 3:} Based on the outcome $\mu$, Bob implements an optimized local unitary on B, $U_\textsc{b}\left(\mu\right)$, to extract previously unavailable energy through B
(See Fig.~\ref{fig:QETprotocol})

%
%

%
%
%

The effect of repeatedly and identically applying the QET protocol to a state $\rho$ can be described by the evolution of the density matrix~\cite{hotta2010energy}:
\begin{equation}
\rho_f=\sum_{\mu=\pm1}U_\textsc{b}(\mu)M_\textsc{a}(\mu)\rho M^{\dagger}_\textsc{a}(\mu)U_\textsc{b}^\dagger(\mu),
\label{eq:rho2}
\end{equation}
where $M_\textsc{a}\left(\mu\right) \: {\rm is}$ the measurement operator of a POVM on A with outcome $\mu$ that commutes with V; and $U_\textsc{b}\left(\mu\right)$ is an informed unitary that maximizes the energy extraction depending on the outcome. Then, the amount of energy extracted locally from B on average is given by 
\begin{equation}-\Delta E_\textsc{b}=-{\rm Tr}\left[\left(H_\textsc{b}+V\right)\rho_f\right]\geq0\label{eq:extracted},
\end{equation} 
since $\left[M_\textsc{a}(\mu),V\right]=\left[M_\textsc{a}(\mu),H_\textsc{b}\right]=0 \: {\rm and}$ given that the expectation value of each term of the Hamiltonian was set to zero
(see Appendix C3 in Ref.~\cite{rodriguez2020novel} and Ref.~\cite{hotta2010energy} for the detailed calculation). It is important to note that if the measurement outcome is not communicated to B (i.e. having $U_\textsc{b} \: {\rm independent}$ ${\rm of} \: \mu$), it is impossible to extract energy from subsystem B on average. This underscores the importance of communication in QET.
 
From Eq.~(\ref{eq:extracted}), the amount of extractable energy is bounded by $0\leq-\Delta E_\textsc{b}\leq - \lambda_{min}$, where $\lambda_{min}$ is the most negative eigenvalue of $H_\textsc{b}+V$. This upper bound is tight when the POVMs are proportional to projective operators. 
In particular, 
for the ground state of the Hamiltonian in Eq.~(\ref{eq:Hamiltonian}), the maximum amount of extractable energy from subsystem B is
\begin{align}
 \displaystyle
    \left(-\Delta E_\textsc{b}\right)_{\rm max}&=-\sqrt{h_\textsc{b}^2+4\kappa^2}+\frac{h_\textsc{b}\left(h_\textsc{a}+h_\textsc{b}\right)+4\kappa^2}{\sqrt{\left(h_\textsc{a}+h_\textsc{b}\right)^2+4\kappa^2}},
    \label{eq:energyextracted}
\end{align}
and can be activated by QET, using projection operators of observable $\sigma_x^\textsc{a}.$ In this case, the average energy injected into A is $E_\textsc{a}=h_\textsc{a}/\sqrt{1+\frac{4\kappa^2}{\left(h_\textsc{a}+h_\textsc{b}\right)^2}}\geq\left(-\Delta E_\textsc{b}\right)_{\rm max}$.
Note that the energy $E_\textsc{a}$ injected into subsystem A remains in that subsystem during the protocol, since the time scale $t_\mu$ for transmitting classical communication from A to B is much shorter than the time scale for energy to propagate from A to B (i.e., $t_\mu\ll t_c\sim1/\kappa$). As a result, the energy extracted corresponds only to the activated energy within the local zero-point energy density.

%


%
%
%

\textit{Experimental implementation.---}To perform the protocol using nuclear magnetic resonance (NMR), we designed a fully unitary QET by introducing an auxiliary system to mediate the measurement of A and transmit information to B, as detailed below. {\color{black}  
The fully unitary version of the QET protocol is equivalent to the minimal QET since the role of a general measurement device can be played by an auxiliary system (An) together with unitary dynamics~\cite{nielsen2002quantum}. This equivalence is proven in Supplemental Material~\cite{supplmat}.
The equivalence 
of a fully unitary QET and the minimal QET has been discussed in Refs.~\cite{funai2017engineering,PhysRevA.93.022308,Verdon2016}.}

\begin{figure}[ht]
\centering
\includegraphics[width=1.\linewidth]{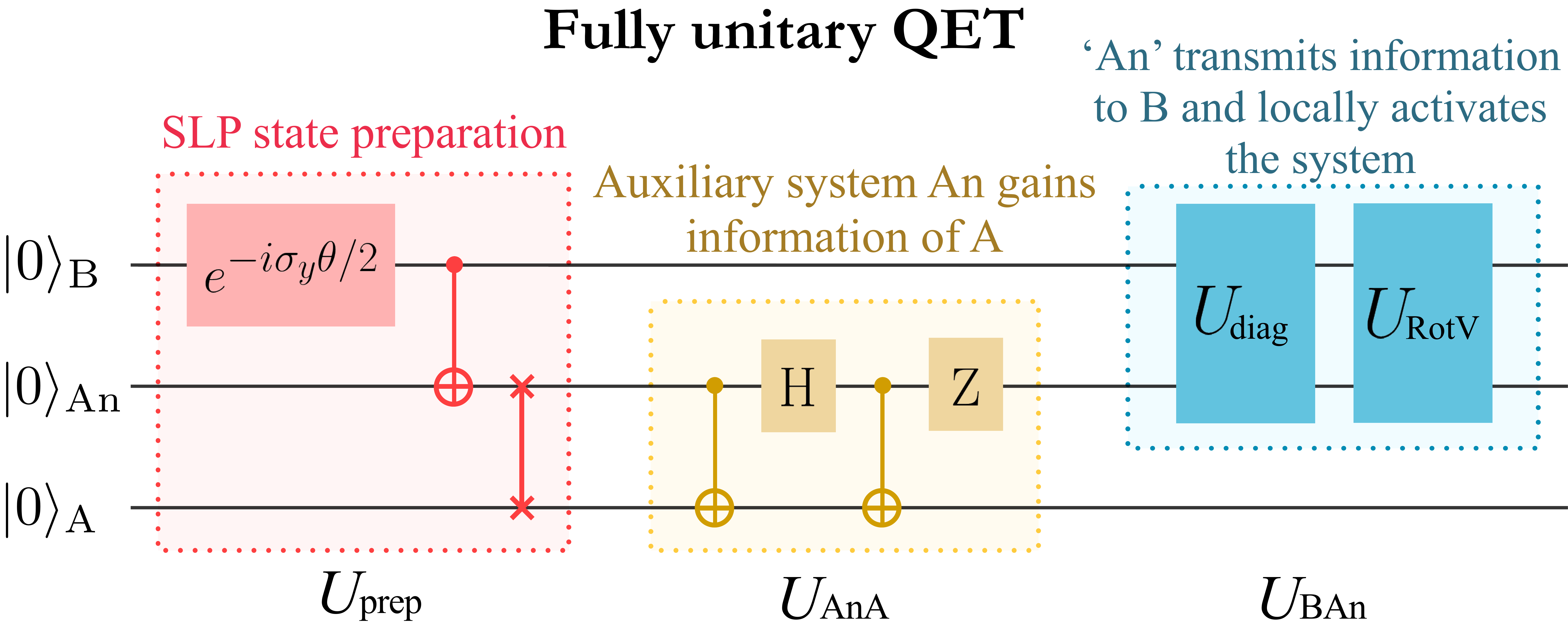}
\caption{{\color{black} Fully unitary QET protocol, performed by adding an auxiliary qubit (An) to mediate the measurement of A and transmit information to B. This circuit was optimized to locally activate the passive state given in Eq.~(\ref{eq:ground}), of Hamiltonian $H\propto -h_\textsc{a}\sigma_z^\textsc{a}-h_\textsc{b}\sigma_z^\textsc{b}+2\kappa\sigma_x^\textsc{a}\sigma_x^\textsc{b}\:\: {\rm and} \:\: H_{\rm An}\propto \sigma_z^{\rm An}$. See Supplemental Material~\cite{supplmat} for the QET protocols' equivalence and details on the experimental implementation.
}}
\label{fig:Protocolc}
\end{figure}

\begin{figure*}[ht]
\centering
\includegraphics[width=1\linewidth]{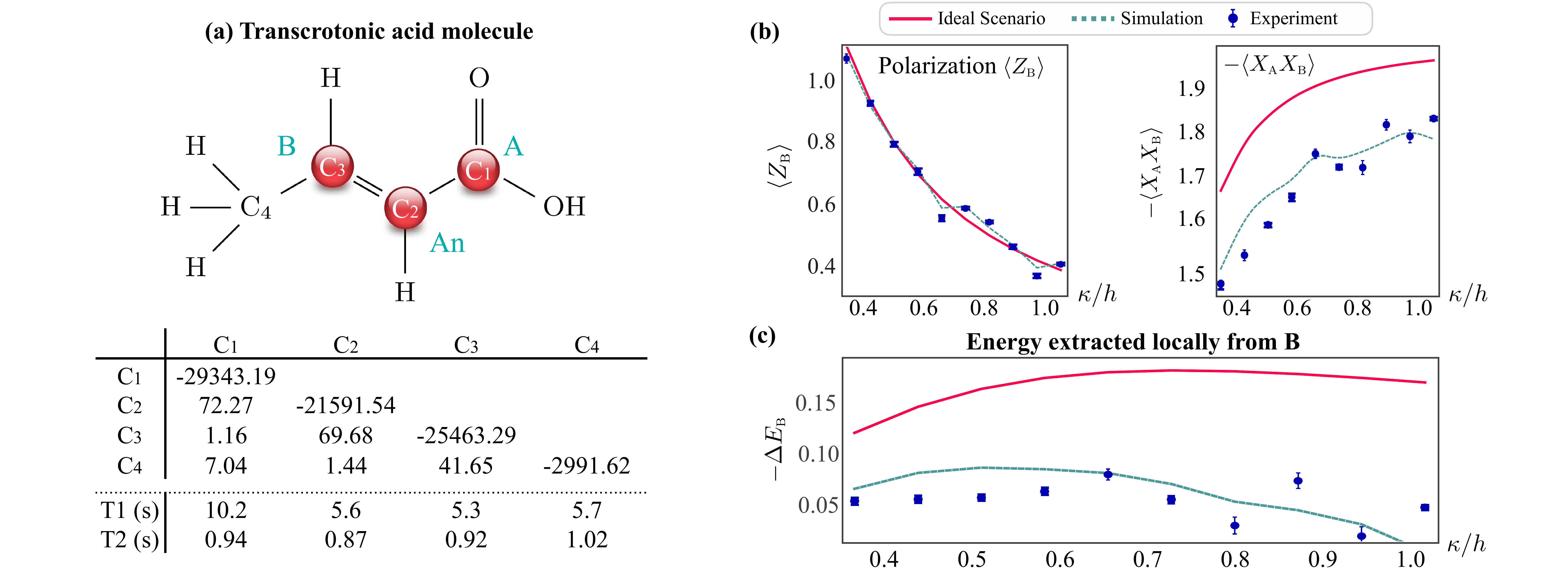}
\caption{\textbf{(a)} System set up in a transcrotonic acid molecule. The table shows the Hamiltonian parameters: the diagonal values correspond to the chemical shifts, and off-diagonal elements correspond to the J-coupling values, all in Hz. {\color{black} The relaxation time scales, $T_1$ and $T_2$ are shown in the bottom.} \textbf{(b)} Experimental results: Expectation values  $-\langle X_\textsc{a}X_\textsc{b}\rangle$ and $\langle Z_\textsc{b}\rangle$ after implementing the protocol. \textbf{(c)} Energy extracted $-\Delta E_\textsc{b}$, plotted against the coupling strength $\kappa/h$ between systems A and B, for $h_\textsc{b}=0.4~h_\textsc{a}$ and fixing $h_\textsc{a}=h=1$.}
\label{fig:Results}  
\end{figure*}

The experiments were performed on a Bruker Avance III 700 MHz NMR spectrometer, using $^{13}\textrm{C}$-${\rm labelled}$ transcrotonic acid dissolved in \mbox{acetone-d$6$} as the sample. The carbons labeled as ${\rm C}_1, \: {\rm C}_2, \: {\rm and} \: {\rm C}_3$ were used as subsystems $\rm B,\: An,\: and\: A,$ respectively [Fig.~\ref{fig:Results}(a)]. The entire experiment took place at an ambient temperature of 298~K. The fully unitary QET protocol consists of the following steps (an overview of the experimental scheme is shown in Fig.~\ref{fig:Protocolc}):
 
\textbf{Step 0:} Preliminary preparation of the SLP state of Eq.~(\ref{eq:ground}): The system starts in the pseudopure state $|000\rangle$~\cite{cory1998nuclear}, followed by a global unitary $U_{prep}.$ The required unitary consists of a rotation $Y(\theta)=e^{-i\sigma_y \theta/2}$ on qubit B, 
followed by a $ \textsc{cnot}$ on A and B (with B as the control). The explicit form of $Y(\theta) \: {\rm is:}$ 
\begin{equation}
    Y\left(\theta\right)=
    -\frac{1}{\sqrt{2}}  \begin{pmatrix}
       F_+ & F_- \\
       -F_- & F_+
    \end{pmatrix},
\end{equation}
$\displaystyle {\rm where} \: F_\pm=\sqrt{1\pm \frac{h_\textsc{a}+h_\textsc{b}}{\sqrt{4\kappa^2+\left(h_\textsc{a}+h_\textsc{b}\right)^2}}}$.

{\color{black} In the experiment, the $\textsc{cnot}$ gate was not directly implemented between subsystems A and B for state preparation, but instead decomposed into two gates acting on subsystems A-An and B-An to improve the state preparation fidelity in our concrete system (see Fig.~\ref{fig:Protocolc}). This is because the J-coupling values of the spin pairs A-An and B-An are higher than the J-coupling of the A-B pair~\cite{fidelityJcouplings}, as shown in Fig.~\ref{fig:Results}. Refer to Supplemental Material~\cite{supplmat} for a more detailed physical explanation.}

 \textbf{Step 1:} Alice gains information about qubit A through an auxiliary qubit An by applying a joint unitary $U_{\rm AnA}$ on both qubits. 
 {\color{black}
 The optimal unitary $U_{\rm AnA}$ corresponds to the one that maximizes the mutual information between A and An, subject to the condition $\left[U_{\rm AnA},V\right]=0$ (so it does not raise the energy of B's surroundings). For a pair of qubits in a product state, an optimal unitary is
\begin{equation}
    \displaystyle
    U_\textsc{\rm AnA}=\frac{1}{\sqrt{2}}
    \begin{pmatrix}
        1 & 0 & 0 & 1 \\ 
        0 & 1 & 1 & 0\\
        0 & -1 & 1 & 0 \\
        -1 & 0 & 0 & 1
    \end{pmatrix}.
\end{equation}
 The explicit gates for this unitary are shown in Fig.~(\ref{fig:Protocolc}).
}
 
%

 \textbf{Step 2:} 
 Alice sends the auxiliary qubit An to Bob in a time $t_\mu$ shorter than the coupling time scale to avoid that energy infused on A propagates to B during the protocol.

\textbf{Step 3.} Finally, Bob implements a joint unitary $U_{\rm BAn}$ on B and An to extract energy from the system by acting locally on B.

For the experiment, we optimized $U_{\rm BAn}$ to achieve the upper bound $\left(-\Delta E_\textsc{b}\right)_{\rm max},$ given by Eq.~(\ref{eq:energyextracted}), obtaining
\begin{align*}
    \displaystyle
    U_{\rm BAn}&=U_{\rm Rot V}U_{\rm diag}, \quad {\rm where}\\
%
    U_{\rm Rot V}&=\frac{1}{\sqrt{2}}
    \begin{pmatrix}
        F_{2_+} & F_{2_-} & 0 & 0 \\ 
        0 & 0 & -F_{2_+}& F_{2_-}\\
        0 & 0 & F_{2_-} & F_{2_+} \\
        -F_{2_-} & F_{2_+} & 0 & 0
    \end{pmatrix}\\
%
    %
    U_{\rm diag}&=\frac{1}{\sqrt{2}}
    \begin{pmatrix}
        0 & F_{+} & F_{-} & 0 \\ 
        F_{-} & 0 & 0 & -F_{+}\\
        F_{+} & 0 & 0 & F_{-} \\
        0 & -F_{-} & F_{+} & 0
    \end{pmatrix},\\
    {\rm with} \quad F_{2_\pm}&=\sqrt{1\pm h_\textsc{b}/\sqrt{h_\textsc{b}^2+4\kappa^2}}.
\end{align*}
 %


The gates for the protocol, shown in Fig.~\ref{fig:Protocolc}, were implemented using GRAPE pulses~\cite{GRAPE} with a slight modification to incorporate the technique described in Ref.~\cite{JPPS}, resulting in the designing of smooth radio frequency (RF) pulses with theoretical fidelity over 0.998 and robust against small imperfections in RF power.


{\color{black} The unitaries $U_{\rm AnA} \: {\rm and} \: U_{\rm BAn}$ were performed in $\backsimeq~10 \: {\rm ms } \: {\rm and}\: 4 \:{\rm ms},$ respectively. Thus, the total time from the beginning of step 1 to the activation of energy was $t_\mu \backsimeq 14 \: {\rm ms}.$ This time fulfills the condition for QET: $t_\mu \ll t_c,$ where $t_c \sim 1/J_\textsc{ab}=(1.16 ~{\rm Hz})^{-1} \: {\rm for}$ the transcrotonic molecule. Namely, $t_\mu$ is much shorter than the time it would take for energy to propagate from A to B.
}

The amount of extracted energy was calculated using the experimental results of the expectation values for the interaction term operator $-\langle X_\textsc{a}X_\textsc{b} \rangle$ and B's local Hamiltonian operator $\langle Z_\textsc{b} \rangle$, which were measured directly at the end of the circuit. The experimental results are plotted in Fig.~\ref{fig:Results}, as a function of the coupling strength between A and B.
The curves for the ideal scenario show the expected outcome of implementing the circuit perfectly, without any decoherence. On the other hand, the curves for the simulation were generated using optimized GRAPE pulses that take decoherence into account. The decoherence simulation assumes that the environment is Markovian, the qubits relax independently, and the dissipator commutes with the total Hamiltonian for the time discretization in GRAPE pulses ($\Delta t = 2~\mu{\rm s}$). These assumptions simplify the implementation of master equations for each time step, the evolution under the propagator of the GRAPE pulses, and the dissipator. All of the experimental results give energy extraction, providing the first evidence of activation of a strong local passive state under experimental conditions. The discrepancy between the simulation and the experiment is due to the decoherence assumptions and the transfer function of the spectrometer, which executes the GRAPE pulses slightly differently than assumed. The error bars shown represent only the statistical error of the experiment.

%
%

%
Numerical tests show that the protocol is stable under some uncertainty in the local Hamiltonians. To study this sensitivity, we added perturbations in the local parts of the Hamiltonian while performing the optimized QET protocol for the non-perturbed case. We considered perturbed local Hamiltonians of the form $h_\nu\propto \left(1+\epsilon\right) h_\nu\sigma_z^\nu+h_\nu f(h_\textsc{a},h_\textsc{b},\kappa)\openone$. We found that if the parameter $\epsilon$ (quantifying the relative difference between the Hamiltonian assumed and the actual Hamiltonian) is small, $\epsilon\leq0.3$, then the relative impact of the error in the implementation of the protocol is neglectable.

\textit{Conclusions.---}We presented the first experimental activation of strong local passive states and the first experimental demonstration of a quantum energy teleportation (QET) protocol proposed by Hotta~\cite{hotta2008protocol,hotta2009quantum}.
Furthermore, our experiment confirms for the first time that the presence of
 entanglement in a ground state allows for local zero-point energy density activation without energy transfer through the system. 

We show experimental energy extraction from a bipartite system, initially in a strong local passive state but activated through local operations and communication, using a quantum energy teleportation protocol. We designed a fully unitary quantum energy teleportation protocol optimized to maximize the energy extraction under the constraints of our experimental setup. The experiment was carried out using nuclear magnetic resonance, demonstrating that the required control for a quantum energy teleportation protocol can be achieved in realistic experimental scenarios.
Furthermore, the optimization of the fully unitary QET demonstrates that the maximum possible amount of activated energy can only be achieved when the measurement device and the measured subsystem gain full mutual information and the measurement outcome is transmitted to the target subsystem.

The QET protocol has the potential to be a valuable tool for a fundamental understanding of quantum thermodynamics and for quantum technologies. On the fundamental side, QET helps understand quantum fluctuations and their role in fundamental scenarios, from quantum field theory in curved spacetimes to quantum thermodynamics: from black hole physics~\cite{braunstein2007quantum,hayden2007black,hosoya2002quantum,hotta2010controlled} to violations of energy conditions in quantum field theory in curved spacetimes~\cite{funai2017engineering}. On the technological side, it has been proposed as a method for improving the purity locally by exploiting interaction-induced correlations in algorithmic cooling protocols~\cite{rodriguez2017correlation}. Another application appears in scenarios where SLP can potentially impose restrictions on thermodynamic tasks and the regime in which some quantum machines can perform, especially those that rely on energy exchange through local quenching and/or pulses that are fast compared to the dynamics of the system, for example in refs.~\cite{gallego2014thermal,perarnau2018strong,gelbwaser2015strongly,newman2017performance}. This experimental demonstration of QET for the first time paves the way for the experimental implementation and exploration of these protocols in controlled quantum systems. 

\begin{acknowledgements}
\textit{Acknowledgments}.---The authors would like to thank John P. S. Peterson for insightful discussions. N.~A.~R.-B. is supported by the Miller Institute for Basic Research in Science at the University of California, Berkeley. E.M-M. is funded by the NSERC Discovery Program, also gratefully acknowledges the funding of the Ontario Early Research Award. R.~L. is supported by a Mike and Ophelia Lazaridis Chair. 
\end{acknowledgements}

\bibliographystyle{apsrev4-1}
\bibliography{references}

\begin{thebibliography}{55}
\bibitem [{zer()}]{zero-pointdensity}%
  \BibitemOpen
\bibfield  {journal} {  }\href@noop {} {\bibinfo  {journal} {{\color{black} In
  general, the entangled ground state of a multipartite quantum system
  possesses regions of positive (and negative) energy density, depending on the
  expectation values of the Hamiltonian terms that have support in that
  particular region}}\ }\BibitemShut {NoStop}%
\bibitem [{fid()}]{fidelityJcouplings}%
  \BibitemOpen
  \href@noop {} {\bibinfo  {journal} {{\color{black} The stronger the coupling
  between spins, the higher the fidelity of a global unitary on both. Direct
  coupling between B and A is smaller and would result in a longer gate time.
  For comparison, A $\textsc{cnot}$ gate between B and A would take roughly 431
  ms while using An to mediate the gate is around 26 ms}}\ }\BibitemShut
  {NoStop}%
\end{thebibliography}

\newpage
\section{Supplemental Material
}

This Supplemental Material presents explicit derivations of how the minimal QET protocol can be reproduced in a fully-unitary version by adding an auxiliary system. It also describes the experimental implementation, including the restrictions in NMR and how they can be overcome to implement the fully-unitary QET protocol.

The material is structured as follows:
Sec.~S1 presents the proof of equivalence between the minimal and the fully-unitary QET protocols.
Sec.~S2 gives the experimental requirements and how using the transcrotonic molecule in NMR can meet these requirements (such as the time scales and the implementation of the circuit for the fully-unitary QET picture).
Finally, sec.~S3 gives the experimental implementation details, including the explicit form of the gates in the circuit and the experimental NMR spectra obtained in the experiment.

\section{S1. Equivalence between the minimal QET and the Fully-Unitary QET}

This section provides a detailed discussion of the equivalence between the minimal QET protocol and the fully-unitary version. The conceptual equivalence and the equivalence in yield of the two versions are connected by the fact that the action of a unitary coupling of a detector and a system, followed by a measurement on the detector, results in a positive operator-valued measure (POVM) on the system~\cite{nielsen2002quantum}. To demonstrate this, let us first review the steps of the minimal QET protocol.\\

\textbf{Minimal QET protocol:}

\begin{enumerate}
    \item The protocol begins with an initial state of the form given by equation~(3) in the manuscript, which is a fully entangled state of the form:
 \begin{align} \label{eq:ground} 
  \ket{g}=\left(F_{+}\ket{00}_\textsc{ab}-F_{_-}\ket{11}_\textsc{ab}\right)/\sqrt{2}.
 \end{align}

    \item Next, in the minimal QET protocol, subsystem A is measured using a POVM with measurement operators $M_\textsc{a}(\mu)$ of outcome $\mu$.
    
    \item The outcome $\mu$ is communicated to B, in a time $t_\mu$ shorter than the coupling time scale to avoid that energy infused on A propagates to B during the protocol. 

    \item Finally, based on the outcome, an informed unitary (which depends on the outcome $\mu$) $U_\textsc{b}(\mu)$ is applied on B, which has a negative energy cost.\\
    
    Thus, the total effect on average of the minimal QET protocol on B is given by
\begin{equation}
\rho_\textsc{b}=\tr{\left[\sum_{\mu}U_\textsc{b}(\mu)M_\textsc{a}(\mu)\ket{g}\bra{g} M^{\dagger}_\textsc{a}(\mu)U_\textsc{b}^\dagger(\mu)
\right]}
\label{minimalQETresult}
\end{equation}

\end{enumerate}
Then, for the fully unitary version, where an auxiliary system mediates the measurement and the classical information channel is replaced with a quantum channel, the steps are as follows:\\

\textbf{Fully-Unitary QET protocol:}

\begin{enumerate}
    \item The protocol begins with the same initial state 
        $\ket{g}=\left(F_{+}\ket{00}_\textsc{ab}-F_{_-}\ket{11}_\textsc{ab}\right)/\sqrt{2}$,
    mentioned in equation~(1),\\

    \item  An auxiliary system `An' is introduced to reproduce the effect of the measurement operators $M_\textsc{a}(\mu)$ with outcomes $\mu =\{0,1\}$. The auxiliary qubit An interacts with subsystem A so that the mutual information between A and An is maximized. This is achieved by implementing the unitary $U_\textsc{\rm AnA}$ given by equation~(8) in the manuscript. It can be verified that the state of An and A will be maximally entangled after implementing $ U_\textsc{\rm AnA}$. Indeed, the unitary $ U_\textsc{\rm AnA}$ maps the computational basis of  An and A to the Bell states basis ($\Phi^{\pm}=\frac{1}{\sqrt{2}}\left(\ket{00}_\textsc{\rm AnA}\pm\ket{11}_\textsc{\rm AnA}\right)$ and $\Psi^{\pm}=\frac{1}{\sqrt{2}}\left(\ket{01}_\textsc{\rm AnA}\pm\ket{10}_\textsc{\rm AnA}\right)$).
After applying the unitary $U_\textsc{\rm AnA}$, the whole system will be in the state:\\

$\ket{\Psi_{_\textsc{\rm BAnA}}}=\frac{1}{\sqrt{2}} \left(F_{+}\ket{0}_\textsc{b}\ket{\Phi^{-}}_\textsc{\rm AnA}-F_{_-}\ket{1}_\textsc{b}\ket{\Psi^{-}}_\textsc{\rm AnA}\right)$

\item The auxiliary qubit An is sent to the side B in a time $t_\mu$ shorter than the coupling time scale to avoid that energy infused on A propagates to B during the protocol. 

\item The information encoded in the new state of the auxiliary system An is transmitted to system B by applying a two-qubit unitary transformation on both systems An and B. This is achieved by implementing a conditional unitary $U_\textsc{b}(\mu)$ on B depending on the state of An, which has the following form:
\begin{equation}
    U_\textsc{\rm BAn}=U_\textsc{b}(0)\otimes \ket{0}\bra{0}_\textsc{\rm An}+U_\textsc{b}(1)\otimes \ket{1}\bra{1}_\textsc{\rm An}
\end{equation}
\end{enumerate}

Once these steps are completed, the state of the subsystem B can be expressed as:
\begin{align*}
    \rho_\textsc{b}&=\tr_\textsc{\rm AnA}\left[U_\textsc{\rm BAn}\ket{\Psi_{_\textsc{\rm BAnA}}}\bra{\Psi_{_\textsc{\rm BAnA}}}  U_\textsc{\rm BAn}^\dagger \right]\\
    &=\tr_\textsc{a}\left[\sum_\mu \bra{\mu}_\textsc{\rm An}U_\textsc{\rm BAn}\ket{\Psi_{_\textsc{\rm BAnA}}}\bra{\Psi_{_\textsc{\rm BAnA}}}  U_\textsc{\rm BAn}^\dagger\ket{\mu}_\textsc{\rm An}\right]
\end{align*}

Then, by replacing $U_\textsc{\rm BAn}$:
\begin{align*}
    \rho_\textsc{b}&=\tr_\textsc{a}\left[\sum_\mu \bra{\mu}_\textsc{\rm An}U_\textsc{b}(\mu)\ket{\Psi_{_\textsc{\rm BAnA}}}\bra{\Psi_{_\textsc{\rm BAnA}}}  U^\dagger_\textsc{b}(\mu)\ket{\mu}_\textsc{\rm An}\right]\\
       &=\tr_\textsc{a}\left[\sum_\mu U_\textsc{b}(\mu)\bra{\mu}_\textsc{\rm An}\ket{\Psi_{_\textsc{\rm BAnA}}}\bra{\Psi_{_\textsc{\rm BAnA}}}  \ket{\mu}_\textsc{\rm An}U^\dagger_\textsc{b}(\mu)\right]
\end{align*}

The term $\bra{\mu}_\textsc{\rm An}\ket{\Psi_{_\textsc{\rm BAnA}}}$, explicitly calculated, for $\mu\in\{0,1\}$, gives:
\begin{align*}
    \bra{\mu}_\textsc{\rm An}\ket{\Psi_{_\textsc{\rm BAnA}}}=\left(\frac{\openone-\mu\sigma_x^\textsc{a}}{\sqrt{2}}\right)\ket{g}
\end{align*}

Then, the state of B can be expressed as
\begin{align*}
    \rho_\textsc{b}=\tr_\textsc{a}\left[\sum_\mu U_\textsc{b}(\mu)\left(\frac{\openone-\mu\sigma_x^\textsc{a}}{\sqrt{2}}\right)\ket{g}\bra{g} \left(\frac{\openone-\mu\sigma_x^\textsc{a}}{\sqrt{2}}\right)U^\dagger_\textsc{b}(\mu)\right]
\end{align*}

From here, by comparing this expression with the one obtained from the minimal QET protocol, equation~(\ref{minimalQETresult}), i.e.
\begin{equation*}
\rho_\textsc{b}=\tr{\left[\sum_{\mu}U_\textsc{b}(\mu)M_\textsc{a}(\mu)\ket{g}\bra{g} M^{\dagger}_\textsc{a}(\mu)U_\textsc{b}^\dagger(\mu)
\right]},
\end{equation*}
the effect of the auxiliary system An on the system has the same effect as the original QET with measuring operators $M_\textsc{a}(\mu)=\left(\openone-\mu\sigma_x^\textsc{a}\right)/\sqrt{2}$.

\section{S2. Experimental requirements for the Fully-Unitary QET}

This section outlines the experimental requirements for successfully implementing the QET protocol, accompanied by a description of the restrictions and challenges that may arise in the experimental setup, and discuss how these can be overcome.

\subsection{Time scales required for the Fully-Unitary QET}

The selection of the appropriate molecule for a QET protocol requires careful consideration of time scales to ensure the successful implementation of the protocol and the prevention of energy transmission through the system. Indeed, the time scales are crucial components of
a QET protocol, in the unitary picture the time scale $t_\mu$ is the time it takes from the first unitary interaction between the auxiliary qubit An with subsystem A ($U_\textsc{\rm AnA}$) and the second unitary interaction with subsystem B ($U_\textsc{\rm BAn}$). This time $t_\mu$ must be small enough so the auxiliary system can effectively move from A to B faster than the energy deposited in A could reach B. Therefore, since the speed at which energy propagates is proportional to the coupling strength $J_\textsc{ab}$ between A and B, the condition $t_\mu\ll1/J_\textsc{ab}$ should be satisfied.


This time scale requirement was satisfied using a transcrotonic molecule for implementation. In the transcrotonic molecule, the direct coupling between A and B is $J_\textsc{ab} =$ 1.16 Hz, while the couplings between A and An (72.27 Hz) and between B and An (69.68 Hz) are stronger. These coupling values allow for the rapid implementation of the unitaries $U_\textsc{\rm AnA}$ and $U_\textsc{\rm BAn}$, ensuring that energy has no time to propagate from A to B during the protocol (see times below). Furthermore, the pulses for these unitaries were programmed to be sent immediately one after the other, with the time between them being only the pulse generator's resolution of nanoseconds ($\ll$ 1/1.16 s). This ensures the time condition required for the QET protocol.

An upper bound for $t_\mu$ is the total time of the implementation of the protocol: 
\begin{equation}
t_\mu <t_{\rm total}=t_{U_\textsc{\rm AnA}}+t_{U_\textsc{\rm AnB}}+t_{\rm pulse}
\end{equation}
where $t_{U_\textsc{\rm AnA}}=1/J_\textsc{\rm AnA}\approx 13.8$ ms, $t_{U_\textsc{\rm AnB}}=1/J_\textsc{\rm AnB}\approx 14.3$ ms, and $t_{pulse}\approx 9.5$ ms. 
To guarantee that energy is not being transmitted from A to B we need that \mbox{$t_\mu\ll t_{\textsc{ab}}=1/J_{\textsc{ab}}\approx 862$ ms}, therefore we do have that
\begin{equation}
    t_\mu<t_{total}\approx 37.6 \text{ ms}\ll t_{\textsc{ab}}\approx 862 \text{ ms}\Rightarrow t_\mu \ll t_{\textsc{ab}}  .
\end{equation}

\subsection{Circuit for the fully-unitary QET picture}
In order to improve the gate fidelity of the QET protocol for our specific molecule, we utilized the auxiliary qubit (An) to facilitate all unitaries in the circuit, including the state preparation stage. This approach is depicted in Fig.3 of the main manuscript. Our results show that this gate's decomposition using An, significantly improves the gate fidelity of the protocol's implementation as detailed below.

In the state preparation stage of the QET protocol, the implementation of the unitary $U_{\rm prep}$ required the use of a $\textsc{cnot}$ gate between qubits A and B. While a direct $\textsc{cnot}$ between these qubits could have been used, the coupling between A and B in our system, with a value of $J_\textsc{ab}$=1.16 Hz, would have resulted in a $\textsc{cnot}$ duration of approximately 431 ms. Instead, we utilized the stronger couplings of 72.27 Hz between A and An and 69.68 Hz between B and An to implement an An-mediated $\textsc{cnot}$ between A and B, resulting in a duration of approximately 26 ms. This approach allows for faster implementation and decreased decoherence, leading to improved fidelity in preparing the A-B pair in the desired state.

The decomposition of the unitary $U_{\rm AnA}$ into gates was designed to reproduce the general unitary presented in equation (8) of the main manuscript, which creates the maximum amount of mutual information between a pair of qubits A and An when they are initially in a product state. In our specific initial state, the first $\textsc{cnot}$ gate of $U_{\rm AnA}$ did not have any effect on the system; however, we retained this $\textsc{cnot}$ gate in the decomposition to accurately reproduce the unitary of equation~(8).

\section{S3. Experimental implementation}
\textit{State preparation.---}
The Hamiltonian (in $\hslash$ units) for a $n$-qubit NMR system can be written as
\begin{eqnarray}
H = H_Z + H_J~.
\end{eqnarray}
Here $\displaystyle H_Z =  2 \pi \sum_{j}\omega_j I_z^j $ is the Zeeman Hamiltonian, characterized by the Larmor frequencies $\omega_j$, and $\displaystyle H_J = 2\pi \sum_{j,k}^{j<k} J_{jk} \vec{I}^j \cdot \vec{I}^k$ is the indirect spin-spin coupling Hamiltonian, with $J_{jk}$ values given in the main text Fig. 4(a).  $\vec{I}^j = I^j_x+I^j_y+I^j_z$, where $I^j_{x,y,z}$ is the pauli-$x,y,z$ matrix for $j^{th}$ qubit divided by 2 \cite{levitt2013spin}.

In thermal equilibrium at room temperature, $kT$ is much larger than the
Zeeman energy splittings. So the density matrix of a $n$-qubit system can
be expanded as~\cite{cory1998nuclear}
\begin{eqnarray}
\rho_{\rm eq } = \frac{1}{2^n} e^{-{H}/{kT}} \approx
                \frac{1}{2^n}(\mathbb{I} + \xi{\overline\rho}_{\rm{eq}}) ~.
\label{rhoeq}
\end{eqnarray}
The identity $\mathbb{I}$ represents a background of a uniformly populated
levels, and for homo-nuclear NMR sample, the traceless part
$\displaystyle {\overline\rho}_{\rm eq} = \sum_j I_z^j $
is known as the deviation density matrix. Only the traceless 
${\overline\rho}_{\rm eq}$ is manipulated by operations in NMR
experiments. In a magnetic field of strength 16.4 T at room
temperature, the small dimensionless number $\xi = \hbar\omega_j / kT
\approx 10^{-5}$.

\begin{figure*}[ht]
\centering
\includegraphics[width=0.9\linewidth]{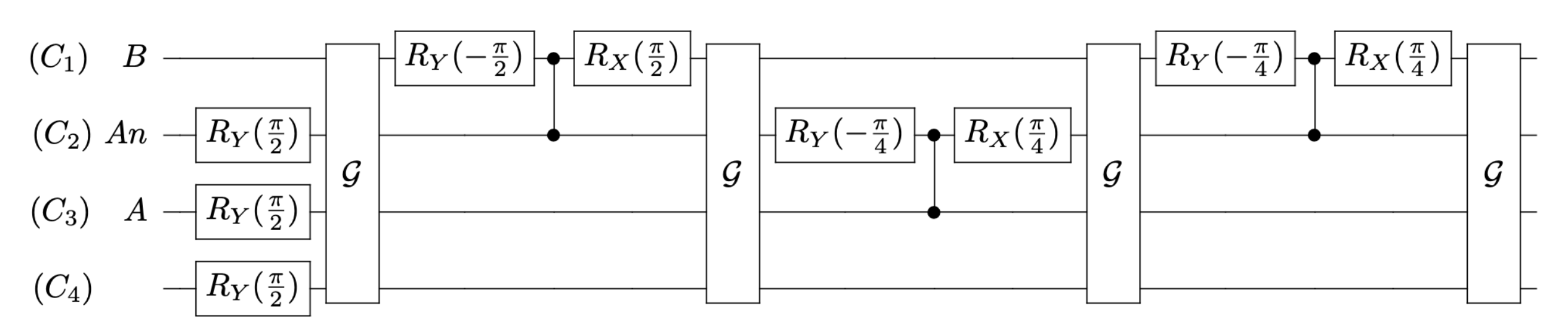}
\caption{Gate sequence 1}
\label{fig:HemantCircuit1}  
\end{figure*}

We prepare a pseudo-pure state from the deviation density matrix using the spatial averaging method \cite{cory1998nuclear}. 
The gate sequence used is depicted in Fig.~\ref{fig:HemantCircuit1},
%
where 
\begin{quantikz}
    & \ctrl{1} & \qw \\
    & \control \qw & \qw \\
\end{quantikz}
is equal to $\displaystyle \exp{(-i\pi I_z^j I_z^k)}$, single qubit gates $R_X(\theta)$ and $R_Y(\theta)$ are the rotation around x,y axis respectively for angle $\theta$, and multi-qubit operation $\mathcal{G}$ represents pulse-field gradient, which destroys all except zeroth order coherence terms \cite{levitt2013spin}. At the end of the sequence, the state is transformed to the pseudo pure state (pps) $\ket{000}\bra{000} \otimes \mathbb{I}$. Note that since we do not use the $4^{th}$ qubit (C$_4$), we have kept it in the maximally mixed state. 

\begin{figure*}[!ht]
    \centering
    \includegraphics[width=0.9\textwidth]{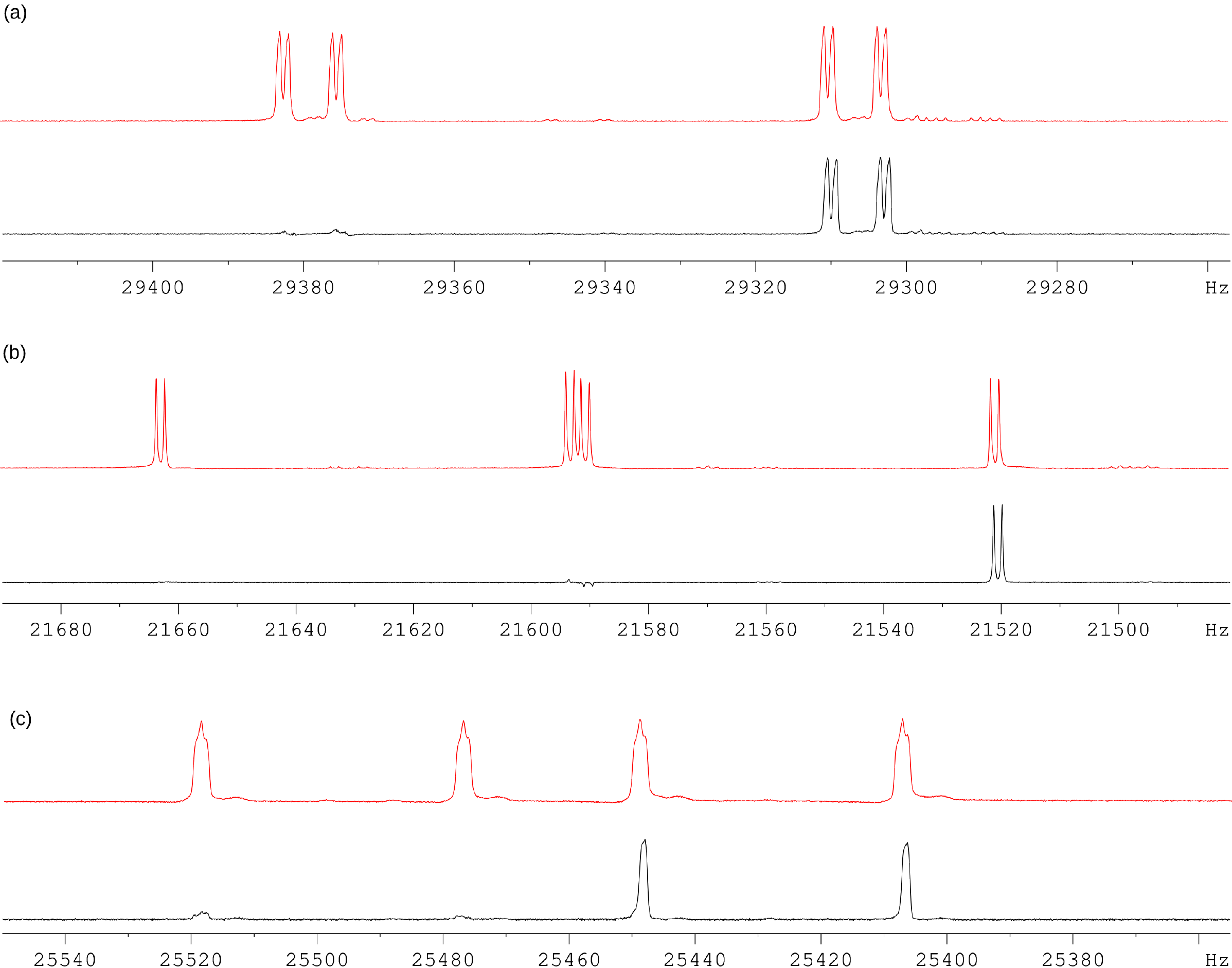}
    \caption{Experimental NMR spectra for pseudo pure state with $4^{th}$ qubit in maximally mixed state. The top (red) spectra is for the thermal state with corresponding pps spectra at the  bottom (black) for (a) qubit 1, (b) qubit 2, and (c) qubit 3. The x axis is the chemical shifts of the qubits (in Hz) while the y axis is in arbitrary units. The y axis for pps were scaled to match the height of thermal state for easier comparison, which were originally roughly $1/8$ the size owing to the signal lost by the application of pulse field gradients and decoherence.}
    \label{fig:nmrspectra}
\end{figure*}

A characteristic of pseudo-pure NMR state is a single peak at its transition frequency. For example, for qubit $1$ if the pps is $|0000\rangle\langle 0000|$, the single peak appears at the frequency $|0\rangle\langle0|(|000\rangle\langle000|)\leftrightarrow |1\rangle\langle1|(|000\rangle\langle000|)$. Since our pps has our 4th qubit in the maximally mixed state, it shows NMR peaks at two transition frequencies; for the previous example it would be at the frequency of $|0\rangle\langle0|(\ket{00}\bra{00}\otimes\mathbb{I})\leftrightarrow |1\rangle\langle1|(\ket{00}\bra{00}\otimes\mathbb{I})$. The experimental NMR spectra for pps is plotted in Figure \ref{fig:nmrspectra} and the fidelity is calculated after tomography to be $0.9996$~\cite{fortunato2002design}.

\begin{figure*}[ht]
\centering
\includegraphics[width=1\linewidth]{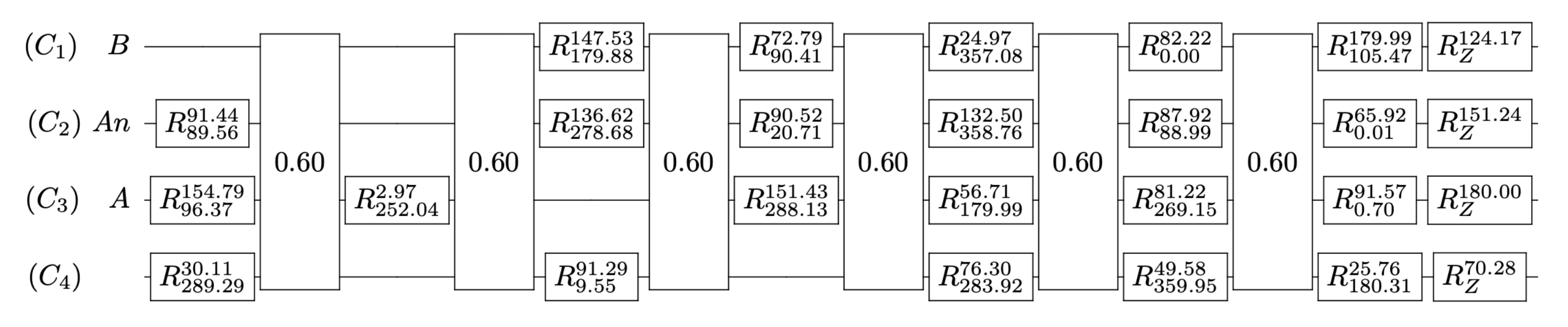}
\caption{Gate sequence 2}
\label{fig:HemantCircuit2}  
\end{figure*}

\textit{Unitary QET protocol implementation.---}
The fully unitary QET protocol was implemented by first decomposing the unitary gates into single qubit gates and evolution under the natural Hamiltonian. For example, the SWAP$_{B,An}$CNOT$_{An,A}$ unitary, which is a part of $U_{\rm prep}$, is decomposed as follows depicted in Fig.~\ref{fig:HemantCircuit2}.
where $ R_{\phi}^{\theta}=  \exp(-i\theta(\cos(\phi)\sigma_x+\sin(\phi)\sigma_y)/2)$, $ R^\theta_Z = \exp(-i\theta\sigma_z/2) $, and multi-qubit gates with a number represent the time of evolution under the natural Hamiltonian in milliseconds. Note that even though the $4^{th}$ qubit doesn't play a part in the protocol, since it is coupled with other spins, we have to apply gates to cancel its evolution. The above sequence actually performs the gate SWAP$_{B,An}$CNOT$_{An,A} \otimes \mathbb{I}$. Similarly, other gates in the protocol ($U_{AnA}$ and $U_{BAn}$) were decomposed, then the GRAPE pulses \cite{GRAPE} were generated for single qubit gates and the evolutions under the natural Hamiltonian were performed with appropriate time delays in the pulse sequence. 

\textit{Measurement.---}
We consider the measurement of first-order coherence terms of a density matrix using nuclear magnetic resonance (NMR) techniques, as described in \cite{levitt2013spin}. The Pauli matrices $X$, $Y$, and $Z$ are used to represent the corresponding operators. For one qubit it's $\langle X \rangle$ and $\langle Y \rangle$, and for two qubits the observable on qubit 1 are $\langle X\mathbb{I} \rangle$, $\langle XZ \rangle$, $\langle Y \mathbb{I} \rangle$, and $\langle YZ \rangle$, and on qubit 2 are $\langle \mathbb{I}X \rangle$, $\langle ZX \rangle$, $\langle \mathbb{I}Y \rangle$, and $\langle ZY \rangle$. If any other term is needed, we need to rotate it in terms of these observables, for example in two qubits if one wishes to measure $\langle XX \rangle$, a single qubit rotation of angle $\pi/2$ along the y axis on qubit 1 will make the term observable on qubit 2 i.e. $\langle ZX \rangle$. For QET protocol we need to measure the expectation value of operators $Z_B$ and $X_AX_B$; both were measured by applying $R_y(\pi/2)$ rotation on qubit $B$, which translated them to $ \mathbb{I}_A X_B$  and $X_A Z_B$. The values were normalized with respect to the experimentally tomographed pseudo pure state.

\end{document}